\documentclass[%
 preprint,
superscriptaddress,
 amsmath,amssymb,
 aps, physrev,
]{revtex4-2}
\UseRawInputEncoding 

\usepackage{graphicx}% Include figure files
\usepackage{dcolumn}% Align table columns on decimal point
\usepackage{bm}% bold math
\usepackage{multirow}
\usepackage{colortbl}
\usepackage{makeidx}
\usepackage[dvipsnames]{xcolor}

\begin{document}

\title{\textbf{Dual Laser and Shadow Particle Image Velocimetry for Multiscale Fluid-Structure Interactions }}
\author{Adrian Herrera-Amaya}
\affiliation{Center for Fluid Mechanics, School of Engineering, Brown University, Providence, RI, 02912, USA}

\author{Monica M. Wilhelmus}
\email{Contact author: mmwilhelmus@brown.edu}
\affiliation{Center for Fluid Mechanics, School of Engineering, Brown University, Providence, RI, 02912, USA}

\date{\today}% It is always \today, today,

\begin{abstract}
Experimental Fluid-Structure Interaction (FSI) often involves phenomena that occur simultaneously across vastly different spatial and temporal scales. For example, the separation bubble behind a cylinder and the vortex interactions in its wake, or the flow produced by an individual swimmer and the flow generated by the movement of the entire swarm. Imaging-based flow studies of such FSI phenomena are challenging and often lead to separate analyses of different flow regions. This work introduces simultaneous shadow and laser-based Particle Image Velocimetry (PIV) to obtain concurrent near-wall and far-field flow velocity measurements. The experimental setup developed here enables the first simultaneous comparison between shadow- and laser-based PIV, demonstrating that shadow velocimetry can achieve accuracy comparable to laser-based measurements at the macro scale. We then use this new combined flow measurement technique to fully capture the velocity field generated by a shrimp-inspired underwater robot, employing laser PIV to record the flow surrounding the robot and shadow PIV to focus on the actuators, imaging regions that the laser technique cannot resolve due to wall reflections and shadows behind the structures. The simultaneous application of shadow- and laser-based PIV offers a new experimental avenue for gaining insights into the physics of FSI phenomena.
\end{abstract}

\maketitle
\newpage
%\tableofcontents

\section{\label{sec:level1}Introduction}

Particle Image Velocimetry (PIV) is a widely used technique in experimental fluid dynamics, providing a quantitative measure of a fluid flow field \cite{raffel_particle_2018,adrian_twenty_2005}. Flow fields obtained through PIV are critical for gaining insights into the physics of a wide range of fluid-dynamics phenomena, including aerodynamics \cite{gardner_review_2023,elsinga_evaluation_2005}, hydrodynamics \cite{ribeiro_wake-foil_2021,grift_hydrodynamics_2021}, biological flows \cite{costello_hydrodynamics_2021,brindise_multi-modality_2019,zade_experimental_2018}, and complex rheology fluids \cite{ravisankar_hydrodynamic_2022,ibezim_micro-piv_2024}. 
\smallskip

\begin{figure}[hbt!]
\centering
\includegraphics[width=\linewidth]{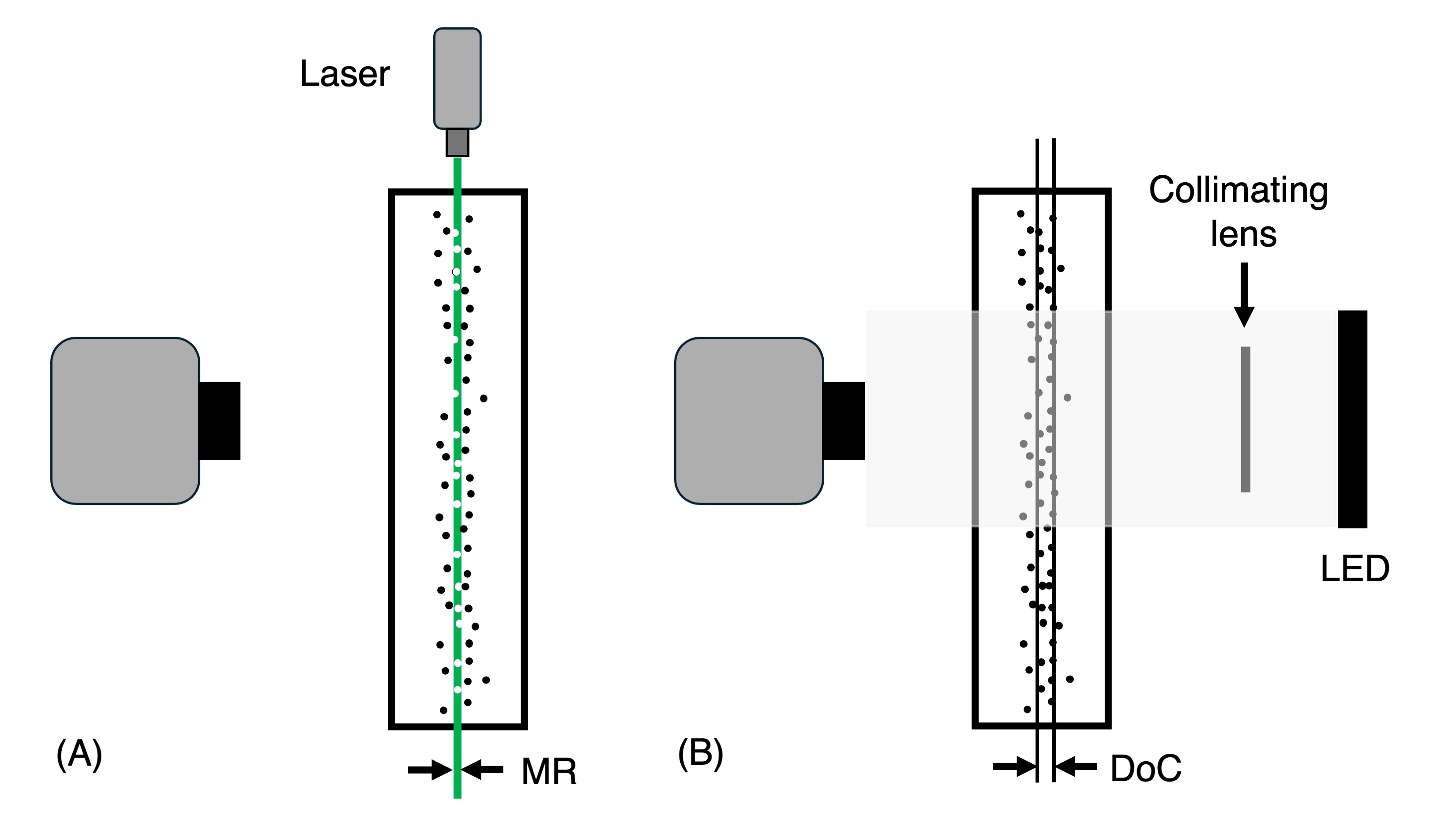}
\caption{Schematic of the experimental setups of two-dimensional two-component image velocimetry acquisition systems. (A) A laser-based system where the thickness of the laser sheet defines the Measurement Region (MR), the camera only sees the particles illuminated in this plane. (B) A shadow-based system where the thickness of the measurement region is defined by the Depth of Correlation (DoC). The DoC is the thickness of the imaged "plane," and it is a property of the optics and tracer particle size.}
\label{fig:PIVPSV}
\end{figure}

In the most common PIV setup, flow fields are measured by estimating the motion of tracer particles illuminated by a laser, using cross-correlation between consecutive time frames (Figure \ref{fig:PIVPSV}A). Laser light can provide high spatial and temporal resolution and a Field of View (FoV) that can span up to a couple of meters \cite{parikh_lego_2023,gunnarson_surfing_2025}. However, laser-based PIV faces considerable challenges when applied to FSI phenomena involving flow measurements around solid objects, such as light scattering and shadow casting. Refractive index matching materials \cite{su_biodegradable_2025,jassal_particle_2025} or mirrors to redirect the light \cite{lucas_pressure-based_2017} work well only for certain combinations of solid-fluid materials or for single solid-object scenarios, making laser-free methods highly appealing for FSI applications.
\smallskip

Shadow-based PIV has been proposed as an alternative to laser illumination for FSI problems \cite{estevadeordal_piv_2005}. Instead of a laser sheet, shadow-based velocimetry employs backlight LED illumination and uses the system optics to isolate tracer motion within a measurement plane (Figure \ref{fig:PIVPSV}B) also known as Depth of Correlation (DoC). For a collimated light source, the DoC is a function of the system optics and the tracer particle size \cite{olsen_out--focus_2000,truong_effect_nodate} and can be estimated by 

\begin{equation}
DoC =2 \sqrt{\frac{1-\sqrt{\epsilon}}{\sqrt{\epsilon}}\left[  \left(\frac{\eta}{2NA}\right)^{2} d_p^{2} + 5.95 \frac{(M+1)^{2} \lambda^{2} \left( \frac{\eta}{2NA} \right)^{4} }{M^{2}}  \right]}\;,
\label{eq:doc}
\end{equation}
\noindent
where $\epsilon$ is a threshold parameter normally set to 0.01 \cite{olsen_out--focus_2000}, $\eta$ is the refractive index of the material between the lens and the specimen, NA is the numerical aperture, $d_p$ is the tracer particle diameter, $M$ is the lens magnification, and $\lambda$ is the illumination wavelength. The PSV technique avoids reflections and shadows when backlighting is feasible, but its FoV is limited. Most shadow-based PIV targets microscale applications, where small volumes, high magnifications, and collimated light result in narrow measurement regions and sharp images \cite{khodaparast_micro_2013,barnkob_general_2015,gemmell_new_2014}. At the macroscale, particle selection, optics, and diffused illumination due to high particle density in large experimental volumes limit the field of view \cite{herrera-amaya_propulsive_2024,hessenkemper_particle_2018,truong_multiplane_2018}, restricting its widespread use despite numerous advantages over traditional laser-based PIV.
\smallskip 

Instead of using shadow-based PIV as an alternative to near-wall flow measurements, we combine it with laser illumination to capture velocity fields more comprehensively, leveraging the strengths of both techniques. Inspired by the logic of simultaneous Planar Laser-Induced Fluorescence (PLIF) and Particle Image Velocimetry (PIV), where a single laser illuminates tracer particles and fluorescent dye and a dichroic mirror separates the wavelengths into two cameras \cite{hu_analysis_2004,naumann_improving_2026}, we employed two light sources -- an infrared laser and a white LED -- to record laser scattering from tracer particles with one camera and shadows cast by the particles with the second camera (Figure \ref{fig:EXPset}).

For the first time, we simultaneously compare laser and shadow-based PIV by recording the motion of a vortex ring within the same FoV in both cameras, showing that macro-scale shadow PIV achieves accuracy comparable to its laser counterpart. Previous comparisons between the techniques have been done by changing the setup from laser to shadow-based PIV and repeating the experiment \cite{jassal_particle_2025, estevadeordal_piv_2005}. We then applied the experimental technique to a multiscale FSI application in which a large field of view is required to record the flow surrounding a shrimp-inspired robot (laser-PIV), and a near-wall approach (shadow-PIV) is needed to capture the flow between the actuators. The resulting velocity field demonstrates the potential of the combined technique to improve image-based flow measurements for a variety of fluid dynamics phenomena, such as bubbly flows \cite{ravisankar_elastic_2025}, bioinspired robotics \cite{peterman_encoding_2024,oliveira_santos_pleobot_2023}, boundary layers \cite{wen_biomimetic_2014}, swarm dynamics \cite{wilhelmus_observations_2014,mohebbi_measurements_2024}, particle-laden turbulence \cite{byron_shape-dependence_2015}, and other FSI phenomena that require the simultaneous study of multiple scales.

\section{Methods}

\begin{figure}[hbt!]
\centering
\includegraphics[width=\linewidth]{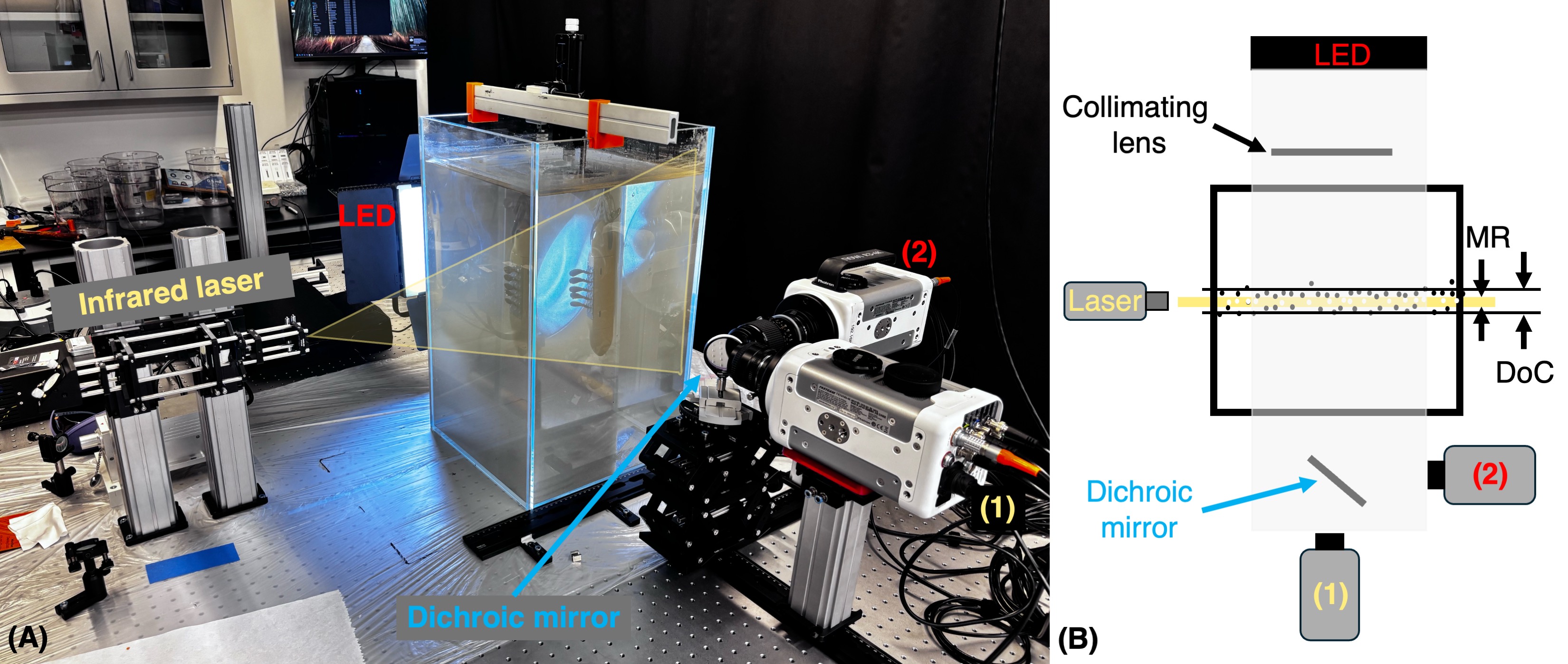}
\caption{Dual-shadow and laser-based experimental setup, (A) a photo showing the shrimp-inspired robot held in place by two linear translation stages (top of the tank) and (B) a top-view schematic showing the arrangement of components. The infrared laser sheet illuminates the experimental tank from the left (as indicated by the yellow schematic), while the LED illuminates the tank from the back. The white LED light passes through the tank and is reflected by the dichroic mirror to camera number two, where the tracer shadows in the DoC are recorded. Meanwhile, laser light scattered by the particles in the MR passes through the dichroic mirror and is recorded by camera number one.}
\label{fig:EXPset}
\end{figure}

To acquire synchronous laser and shadow-based flow imaging, the experimental facility consisted of two illumination sources: an infrared continuous laser ($808 \pm 3$ nm, Opto Engine MDL-N-808-10W) and a 60 W white LED panel (430-450 nm, LituFoto R60). Both sources illuminated an experimental tank measuring 35 x 35 x 60 cm$^3$ from the side and back, respectively (Figure \ref{fig:EXPset}). The tank was filled with a glycerol-water mixture ($\rho_{f} = 1114 \ $kg/m$^{3}$, $\mu = 4.814 \times 10^{-3} \ $Ns/m$^{2}$) and 100 $\mu$m hollow glass silver-coated tracer particles ($\rho_{p} = 1100 \ $kg/m$^3$, Potters Industries). The infrared laser created a 1.25 mm-thick laser sheet to illuminate the tank; the scattered light from tracer particles passed through a dichroic mirror (760 nm cut-on, DMLP760T, Thorlabs) and was recorded by a high-speed camera. Simultaneously, the white LED light was collimated by a 12-inch Fresnel lens placed between the LED and the tank, passed through the tank, reflected off the dichroic mirror with a 45$^\circ$ angle of incidence, and the shadows of tracer particles were captured by a second high-speed camera. Both cameras (Nova R3-4K, Photron) recorded at 500 frames per second, with a shutter speed of $1/2000$ s and a resolution of 4096 x 2304 px. We conducted two experiments: one to compare the performance of shadow-based PIV with that of more traditional laser-based PIV, and another to demonstrate the multiscale capabilities of our new technique.

\begin{figure}[hbt!]
\centering
\includegraphics[scale=0.1]{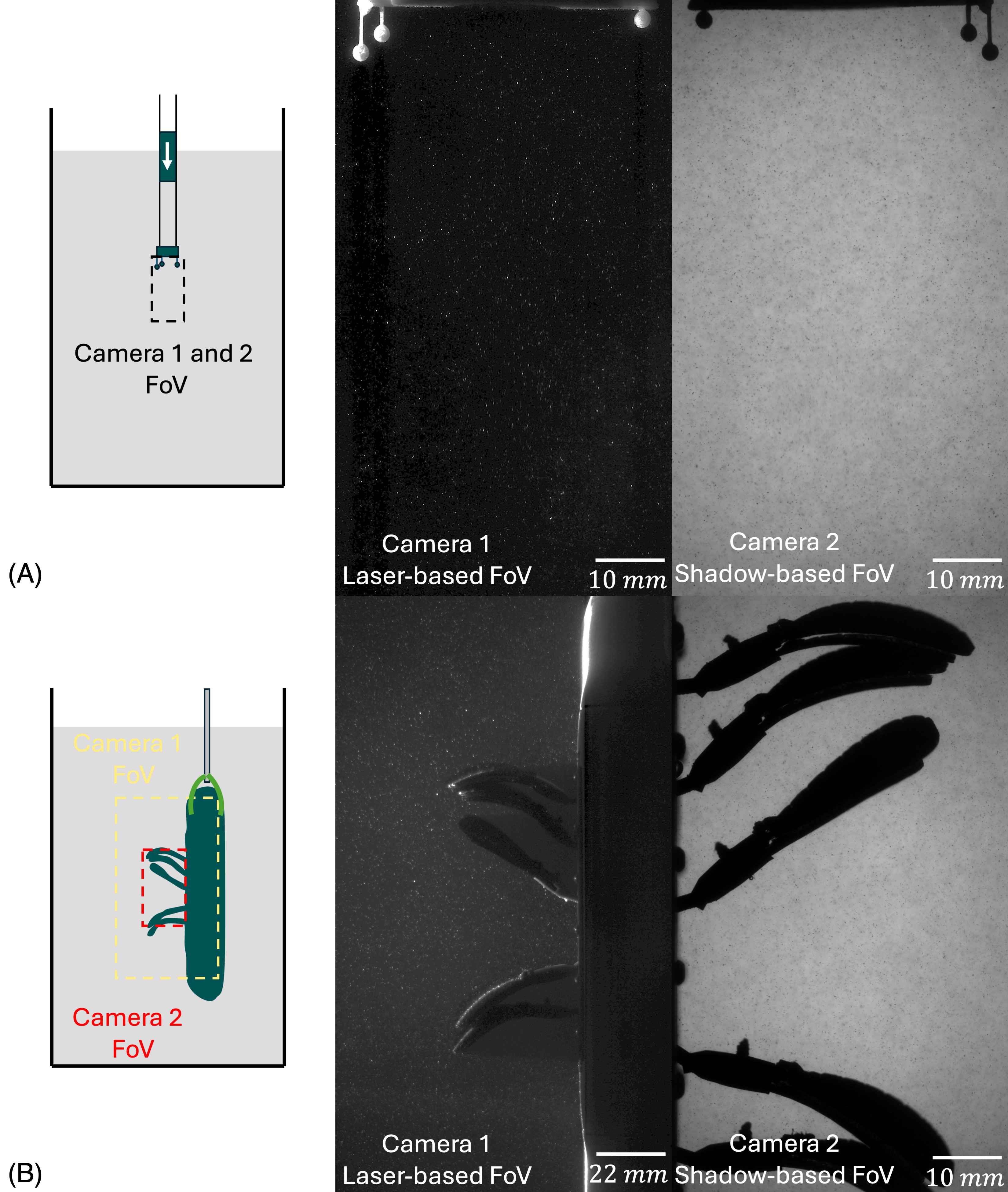}
\caption{Schematics and representative images from the laser and shadow-based recordings; images are mirrored left to right due to the use of a dichroic mirror. (A) Performance comparison vortex ring experiment, and (B) the multiscale shrimp-inspired robot experiment.}
\label{fig:PerfMulti}
\end{figure}

In the first experiment, both cameras filmed the end of a piston-cylinder setup at the center of the tank (Figure \ref{fig:PerfMulti}A), with a field of view measuring $86.23 \times 48.50 \ $mm$^2$ and using 105 mm lenses (Micro- NIKKOR, Nikon). The piston fell under gravity, creating a vortex ring that traveled through the field of view. Laser and shadow-based recordings are inverted; laser illumination produces bright particles on a black background, while LED illumination produces dark particles on a bright background (Figure \ref{fig:PerfMulti}). It is recommended to invert shadow-based recordings so they can be processed by traditional cross-correlation PIV software developed for bright particles on a dark background. When using shadow-based PIV at macro scale, several factors produce larger noise-to-signal ratios than those found in micro PIV. In particular, given the size of the experimental volume and the particle density, light becomes diffused, reducing the sharp shadows observed at micro scales \cite{khodaparast_micro_2013,gemmell_new_2014}. In this experiment, we compare commonly used preprocessing techniques used at micro scales with alternatives that expand the use of shadow-based PIV to larger scales. 

Images from the laser-recording camera were preprocessed by applying an intensity high-pass filter (kernel size $15 \ $px) and an intensity capping operation. Vector computations were carried out using the MATLAB-based tool PIVLab \cite{thielicke_pivlab_2014} (two-pass iteration with $256 \times 256 \ $px and $128 \times 128 \ $px subwindows, 50\% overlap). Finally, all velocity vector fields were postprocessed with a standard deviation filter ($n = 8$) and a local median filter (threshold of 3). Before comparing the velocity fields, we ensured accurate alignment by transforming the shadow-based vectors into the laser-based frame of reference using the positions of the three circular targets placed at the piston exit (Figure \ref{fig:PerfMulti}A).

To compare the similarity between the shadow- and laser-based PIV vector fields, we used the modified Combined Magnitude and Relevance Index (mCMRI) \cite{nowruzi_numeric_2025}. The discrete velocity fields consist of velocity vectors $\boldsymbol{q}$ with components $(u,v)$ associated with a data point $(x_i,y_j)$. The vector magnitude is defined as the $\boldsymbol{L}^2$ norm $\| \bm{q} \|_{2}$. The subscripts ``L'' and ``S'' refer to the laser- and shadow-based PIV velocity fields. To compare the velocity fields, we need to consider the orientation and magnitude of each individual vector. To locally compare the alignment between vector fields, we can use the Local Structural Index \cite{zhao_multi-plane_2019}

\begin{equation}
LSI(x_{i},y_{j}) = \frac{\boldsymbol{q}_{L}(x_{i},y_{j}) \cdot \boldsymbol{q}_{S}(x_{i},y_{j})}{\| \bm{q}_{L}(x_{i},y_{j}) \|_{2} \| \bm{q}_{S}(x_{i},y_{j}) \|_{2}}
\end{equation}

However, the LSI is highly sensitive to low-velocity regions because its denominator includes local vector magnitudes. To mitigate this sensitivity to low-velocity regions, the Weighted Relevance Index (WRI) was introduced by \cite{willman_quantitative_2020} as

\begin{equation}
WRI(x_{i},y_{j}) = \left(\frac{1-LSI(x_{i},y_{j})}{2}\right) \times \left(\frac{\| \bm{q}_{L}(x_{i},y_{j}) \|_{2} \| \bm{q}_{S}(x_{i},y_{j}) \|_{2}}{\text{M}(\bm{Q}_{L})\text{M}(\bm{Q}_{S})}\right)
%WRI(x_{i},y_{j}) = \left(\frac{1-LSI(x_{i},y_{j})}{2}\right) \times \left(\frac{\| \bm{q}_{L}(x_{i},y_{j}) \|_{2} \| \bm{q}_{S}(x_{i},y_{j}) \|_{2}}{median(\bm{Q}_{L})median(\bm{Q}_{S})}\right) \;,
\end{equation}

where $\text{M}(\bm{Q}_{L})$ is the median of $\bm{Q}_{L}$, the magnitude field of the vector field L. The approach involves multiplying the LSI equation by a normalization factor that depends on the median velocity magnitude to adjust the LSI value in high- and low-velocity regions.

%where $\bm{Q}_{L}$ represents the magnitude field of the vector field L, and $\text{M}()$ stands for the median.

Following the same logic of using the median to reduce sensitivity, we can quantify the magnitude difference using the Weighted Magnitude Index (WMI) \cite{willman_quantitative_2020}

\begin{equation}
WMI(x_{i},y_{j}) = \frac{|\bm{Q}_{L}(x_{i},y_{j})-\bm{Q}_{S}(x_{i},y_{j})|}{\text{M}(\bm{Q}_{L},\bm{Q}_{S})}
\end{equation}

Finally we can capture both magnitude and alignment dissimilarities, thereby providing a meaningful overall agreement measure between two vector fields by combining WRI and WMI as 
\begin{equation}
mCMRI(x_{i},y_{j}) = \frac{\frac{WMI(x_{i},y_{j})}{max(WMI(x_{i},y_{j}))}+\frac{WRI(x_{i},y_{j})}{max(WRI(x_{i},y_{j}))}} {2}
\label{eq:mCMRI}
\end{equation}

this approach scales the normalized WMI and WRI, ensuring that both metrics contribute equally when averaged, yielding values in the range [0,1], where 0 indicates identical vectors and 1 signifies that the maximum values of WRI and WMI coincide. 

In the second experiment, the cameras recorded the \emph{Nereus} shrimp-inspired robot developed by the Wilhelmus group at Brown University \cite{tack_going_2025} performing metachronal rowing \cite{herrera-amaya_effects_2025} at a beat frequency of 1 Hz. The shadow camera FoV and vector computation parameters remained the same, whereas the laser recording camera was fitted with a 50 mm lens (AF NIKKOR, Nikon), resulting in a FoV of 193.16 $\times$ 108.65  mm, representing a 400\% increase in recording area compared to the shadow-based FoV (Figure \ref{fig:PerfMulti}B). Vector calculations used a three-pass iteration with 256 x 256 px, 128 x 128 px, and 64 x 64 px subwindows and 50\% overlap. To combine the velocity fields from both techniques, we scaled and aligned the shadow-based velocity vector field with the laser-based frame using three recognizable points along the actuators of the robot (Figure \ref{fig:PerfMulti}B). We then interpolated the velocity vectors from the shadow-based measurement into the laser-based grid. If both grids had a velocity vector, the final merged value was taken as the average (Figure \ref{fig:Pressure}). All image preprocessing and vector field postprocessing were conducted with MATLAB (MathWorks, Inc.). To showcase the improvement of the current technique for multiscale FSI phenomena, we computed the pressure field generated by the rowing motion of the robot using the MATLAB-based tool QUEEN 2.0 \cite{dabiri_algorithm_2014}, which has previously been used to calculate pressure fields of rowing appendages but was always limited to small FoV covering just the neighborhood of the appendage \cite{herrera-amaya_propulsive_2024,colin_role_2020}. The QUEEN 2.0 algorithm integrates the Navier-Stokes equations along eight paths emanating from each point in the mesh and terminating at the boundaries of the field. The pressure at each mesh point is computed as the median pressure from the eight integration paths, making the result sensitive to the FoV size. The robot was masked to prevent surface artifacts in the pressure results. Masks were automatically generated by detecting the appendages from the shadow-based recording and merging them with the robot's body from the laser-based camera.

\section{Results}

\subsection{Similarity between shadow and laser-based velocity vector fields }
\label{subsec:similarity}

\begin{figure}[hbt!]
\centering
\includegraphics[width=\linewidth]{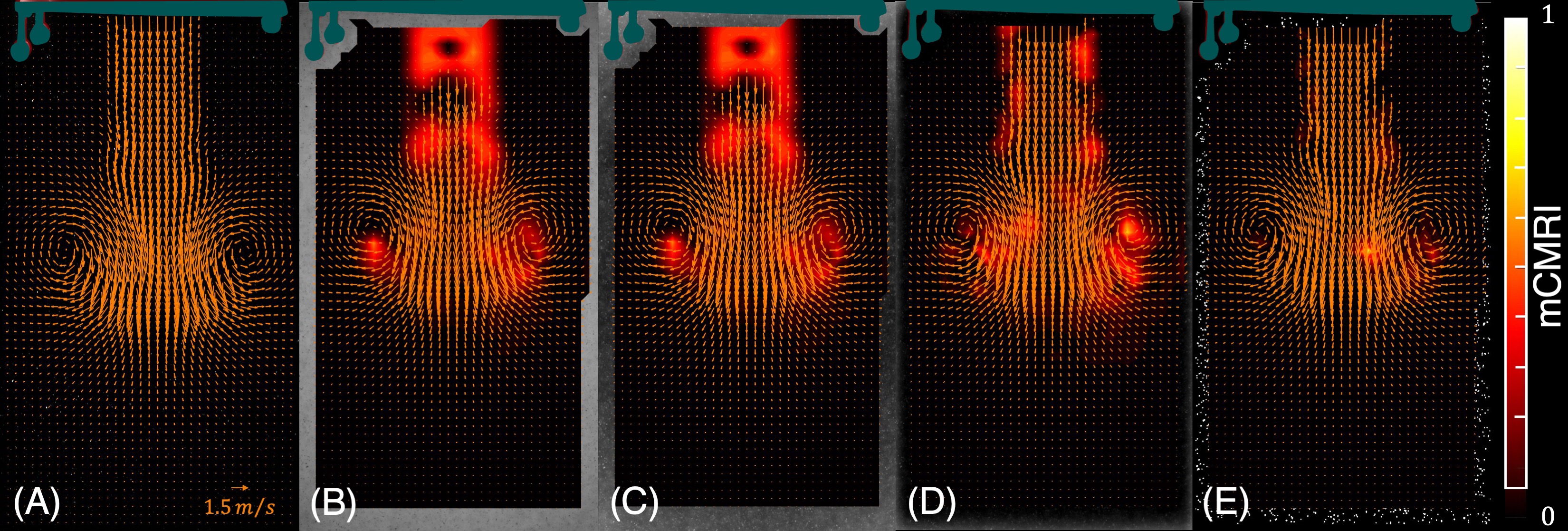}
\caption{Instantaneous modified CMRI comparisons between laser-based PIV (A) and shadow-based PIV using several common image pre-processing techniques: (B) image inversion, (C) intensity capping, (D) high-pass filter, and (E) tracer detection algorithm. Velocity fields computed for each case are plotted with orange arrows. A mCMRI value of 0 indicates identical vectors, while a value of 1 signifies deviations corresponding to maximum WMI and WRI values.}
\label{fig:FieldmCMRI}
\end{figure}

To directly compare the velocity fields obtained by shadow-based PIV with those from the more traditional laser-based PIV, we recorded the motion of a vortex ring traveling down the tank at an average speed ($U_{VR}$) of $0.23 \pm 0.01 \ $m/s and a vortex diameter ($l_{VR}$) of $29.6 \pm 0.4 \ $mm (measured by tracking the vortices as described in section \ref{subsec:Perf}). The Stokes number, defined as the tracer particle relaxation time $\left(\tau_{p} = \tfrac{ d_{p}^2(\rho_{p}-\rho_{f})}{{18\mu}}\right)$ over the flow time scale $\left(\tau_{f} = \tfrac{l_{VR}}{U_{VR}}\right)$ \cite{raffel_particle_2018}, yields $St = - 1.25 \times 10^{-5}$, ensuring that the particles faithfully trace the flow. We compared the shadow-based velocity fields to the laser-based results (Figure \ref{fig:FieldmCMRI}A), applying four different preprocessing techniques (Figure \ref{fig:FieldmCMRI}B-E). 

In the common micro-scale shadow-based PIV setup, where DoC are small and light is often collimated, a simple image inversion or an intensity-capping filter often yields good results \cite{hagsater_investigations_2008,shavit_intensity_2007}. Image inversion computes the complement of the image; for normalized gray images, that is $1 - px \ value$, turning a dark object on a bright background into a bright object on a dark background. Intensity capping sets an upper limit on the grayscale and clips any pixel above it to that value. Bright spots in the area will contribute more to the correlation signal; the intensity-capping filter circumvents this problem. Due to the large FoV in macro-scale shadow-based PIV and the resulting increased noise-to-signal ratio, we also explore the use of a high-pass filter and a tracer detection algorithm developed to regenerate collected images containing only the in-focus tracers \cite{herrera-amaya_focused_2020}. The high-pass filter removes slow, large-scale intensity variations by subtracting a blurred (low-pass) version of the image from the original, thereby suppressing low-frequency content \cite{thielicke_pivlab_2014}. The tracer detection algorithm identifies each focused particle using a series of image filters and morphological operations, generating a new binary image containing only the focused particles \cite{herrera-amaya_adrian_software_2020}.

To compare the velocity fields from the two techniques, we set the optics of the shadow-based PIV to the smallest possible DoC in our experimental setup. With the f-stop ($NA = \frac{1}{2f_{stop}}$) set to 2.8 and a magnification of 0.22, the resulting DoC is $1.7 \ $mm (Eq.\ref{eq:doc}). The modified CMRI (Eq. \ref{eq:mCMRI}) shows that both a simple image inversion (Figure \ref{fig:FieldmCMRI}B) and an intensity capping filter (Figure \ref{fig:FieldmCMRI}C) yield poor matches to the laser-based result (Figure \ref{fig:FieldmCMRI}A). We observe marked differences in both the magnitude and orientation of vectors in the jet and vortex cores. The high-pass filter considerably reduces differences in the jet area while retaining some mismatch in the vortex core region (Figure \ref{fig:FieldmCMRI}D). Finally, the tracer detection algorithm shows better agreement with the laser-based velocity field, reducing vector-field dissimilarity to a small region near the vortex cores (Figure \ref{fig:FieldmCMRI}E).

\begin{figure}[hbt!]
\centering
\includegraphics[width=\linewidth]{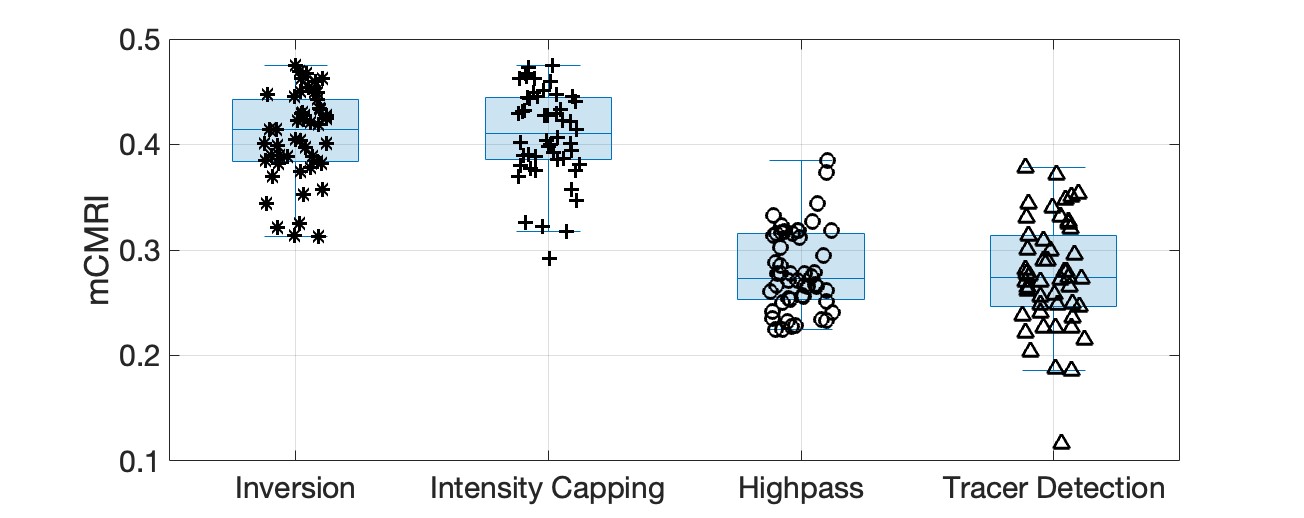}
\caption{Pointwise comparison of laser-based and shadow-based velocity fields across several common image pre-processing techniques. Data points show the 98th percentile of mCMRI at each recorded time step along the vortex ring trajectory $(n = 50)$.}
\label{fig:PCTCMRI}
\end{figure}

To compare the entire sequence of the vortex ring as it descends (50 frames), we used the 98th percentile of the mCMRI field for a pointwise comparison between the shadow-based velocity fields and the laser-based measurements (Figure \ref{fig:PCTCMRI}). The median values for both the high-pass and the tracer detection algorithm show that $98 \%$ of the data is below $mCMRI = 0.27$, while simple inversion and intensity capping yield values under $0.4$. Both the fieldwise (Figure \ref{fig:FieldmCMRI}) and pointwise comparisons indicate that the high-pass and tracer detection algorithms perform better at the macro-scale than the common preprocessing techniques used in shadow-based PIV for micro-scale experiments.

\subsection{Performance comparison of shadow and laser-based PIV }
\label{subsec:Perf}

\begin{figure}[hbt!]
\centering
\includegraphics[width=\linewidth]{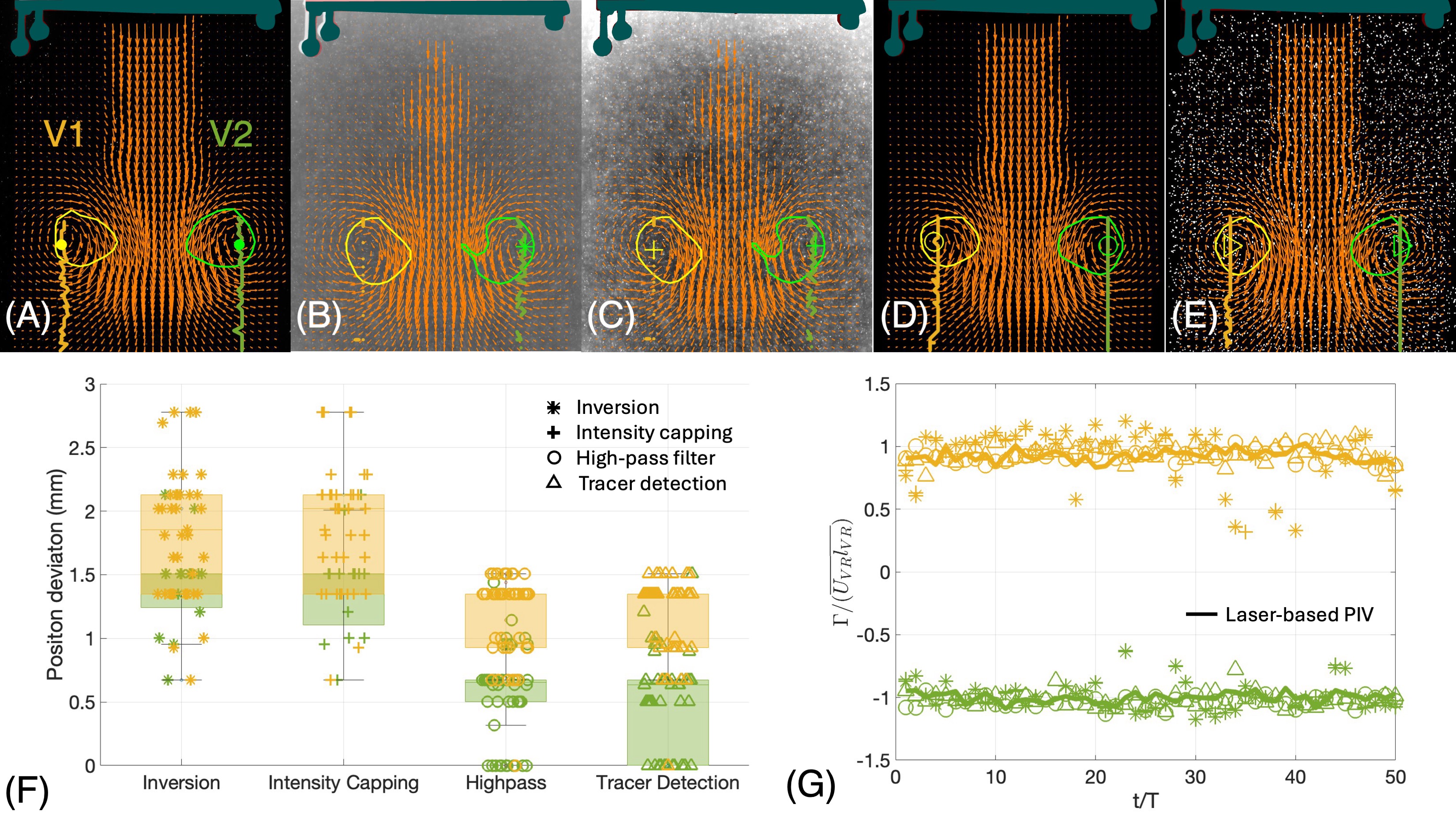}
\caption{Comparison of vortex ring metrics between laser- and shadow-based PIV. Velocity fields show the vortex trajectories identified by the $\Gamma_1(\Gamma_2)$ criterion for the right and left vortices (yellow and green). Dots indicate vortex centers at equal time intervals along the vortex ring trajectory. The vortex cores identified using the $Q$-criterion are shown for the current time interval by closed solid lines. (A) Laser-based PIV velocity field, and the different image pre-processing techniques applied to the shadow-based PIV data: (B) Image inversion, (C) Intensity capping, (D) High-pass filter, and (E) Tracer detection algorithm. (F) Position deviation of the vortex centers compared to the laser-based PIV data. (G) Normalized circulation of the vortex cores V1 and V2.}
\label{fig:VortexD}
\end{figure}

For a more physical comparison, we identified the vortex ring traveling downward by using the $\Gamma_1$ and $\Gamma_2$ criteria \cite{graftieaux_combining_2001} to locate the centers of the left and right vortices (V1 and V2) and the $Q$ criterion \cite{hunt_eddies_1988} to identify the vortex core (Figure \ref{fig:VortexD}A-E), following the framework described in \cite{huang_detection_2015}. Consistent with the mCMRI index results, single inversion and the intensity cap filter preprocessing for shadow-based velocimetry yielded the worst performance. The simple inversion had a success rate in finding the vortex center of $30 \%$ for V1 and $78 \%$ for V2 (Figure \ref{fig:VortexD}B), while the intensity cap filter showed $32 \%$ for V1 and $76 \%$ for V2 (Figure \ref{fig:VortexD}C). In terms of precision in locating the vortex center compared to the laser-based technique (Figure \ref{fig:VortexD}A), both showed position deviations as high as $2.77 \ $mm, equivalent to $9.78\%$ of the vortex ring diameter (Figure \ref{fig:VortexD}F). Both the highpass filter and the tracer detection algorithm image preprocessing achieved a $100 \%$ success rate in locating the vortex center (Figure \ref{fig:VortexD}D-E) and maintained precision under $1.50 \ $mm or $5.3\%$ of the vortex ring diameter (Figure \ref{fig:VortexD}F).

The higher noise-to-signal ratio in the simple inversion and intensity capping results in a highly deformed vortex core boundary (Figure \ref{fig:VortexD}B-C), whereas the high-pass filter and tracer detection algorithm show vortex core boundaries that more closely resemble the smooth circular cores identified in the laser-based recordings (Figure \ref{fig:VortexD}D-E). To quantify the effect of the core distortions, we compare the normalized circulation along the core boundaries for the entire vortex ring trajectory (Figure \ref{fig:VortexD}G). The simple inversion and intensity capping are less accurate at capturing the evolution of the vortex circulation than the high-pass filter and tracer detection, relative to the laser-based data. On the one hand, the normalized Root Mean Square (nRMS) of the differences between curves places the agreement of the simple inversion preprocessing within $24.74\%$ of the mean of the laser-based data for V1 and $11.68\%$ for V2, and the intensity capping within $26.14\%$ for V1 and $11.79\%$ for V2. On the other hand, the high-pass filter has an agreement within $6.41\%$ for V1 and $9.39\%$ for V2, and the tracer detection algorithm within $6.47\%$ for V1 and $6.28\%$ for V2. When we take into account the three performance indicators we calculated, namely the accuracy in detecting the vortices, the position deviation of the vortex center, and the nRMS of the vortex circulation evolution, it is clear that the high-pass filter and the tracer detection algorithm are the ones capable of achieving the data quality that modelers need to validate numerical models or to train AI approaches \cite{zhang_reconstructing_2025,brunton_machine_2020}.

\subsection{Depth of correlation impact on macro-scale shadow-based PIV }

When setting up a shadow-based PIV experiment, the goal is to keep the DoC as small as possible to ensure imaging only particles on the focal plane. For our new dual-technique approach, the DoC should match the laser sheet thickness to image the same volume with both techniques; however, this is not always possible. For a specific fluid medium and light source, the DoC is a function of the numerical aperture, the tracer particle diameter, and the magnification --- the lower these parameters are, the smaller the resulting DoC (Eq. \ref{eq:doc}). Applying this technique to a large FoV, however, requires low magnifications, and particle size is limited by the camera sensor resolution. For the vortex ring experiment, the lowest possible $f_{stop}=2.8$ results in a DoC of $1.7 \ $mm, $36 \%$ larger than the laser sheet measurement region of $1.25 \ $mm.  

\begin{figure}[hbt!]
\centering
\includegraphics[scale=0.37]{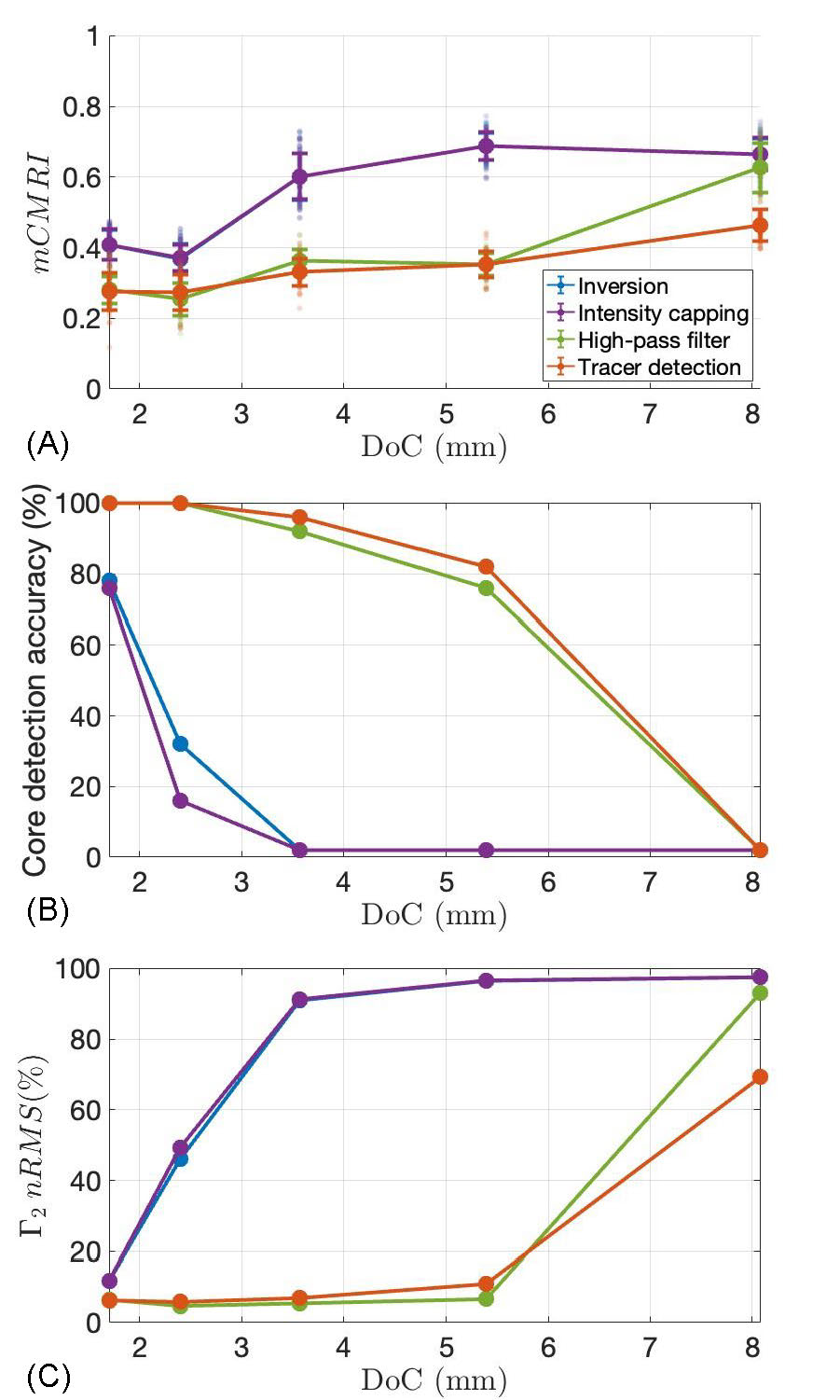}
\caption{Vector field similarity and performance comparison between laser- and shadow-based PIV across different depths of correlation and preprocessing techniques. (A) Pointwise comparisons of the 98th percentile of the mCMRI. Filled circles denote the mean mCMRI value for the vortex-traveling sequence; translucent points show each time-interval comparison; and error bars represent the standard deviation $(n=50)$. (B) Accuracy in detecting the vortex V2 center. (C) Mean normalized root mean square of the differences between vortex core circulation curves during the vortex V2 trajectory.}
\label{fig:DoC}
\end{figure}

We investigated the effect of the DoC by using all the f-stop values of the $105 \ $mm lenses ($2.8, 4, 5.6, 8, 11$). The 98th percentile of the mCMRI shows that preprocessing techniques such as the high-pass filter and the tracer detection algorithm provide greater flexibility, yielding better agreement with the laser-based data than simple inversion or intensity capping. The 98th percentile mCMRI remains under $0.4$ even for a DoC $331.2 \%$ larger than the laser sheet thickness (Figure \ref{fig:DoC}A). In terms of performance, the accuracy in finding the vortex centers and the vortex circulation nRMS worsen significantly for the simple inversion or intensity capping filters, leading to unacceptable noise for any DoC larger than the minimum (Figure \ref{fig:DoC}B-C). Both the high-pass filter and the tracer detection algorithm retain an accuracy finding the vortex center over $92 \%$ and a circulation nRMS within $6.94 \%$ of the mean laser-based mesurements for DoC up to $3.57 \ $mm, $185.6 \%$ larger than the laser-based PIV measurement region.

\subsection{Multiscale shadow and laser-based PIV}

\begin{figure}[hbt!]
\centering
\includegraphics[width=\linewidth]{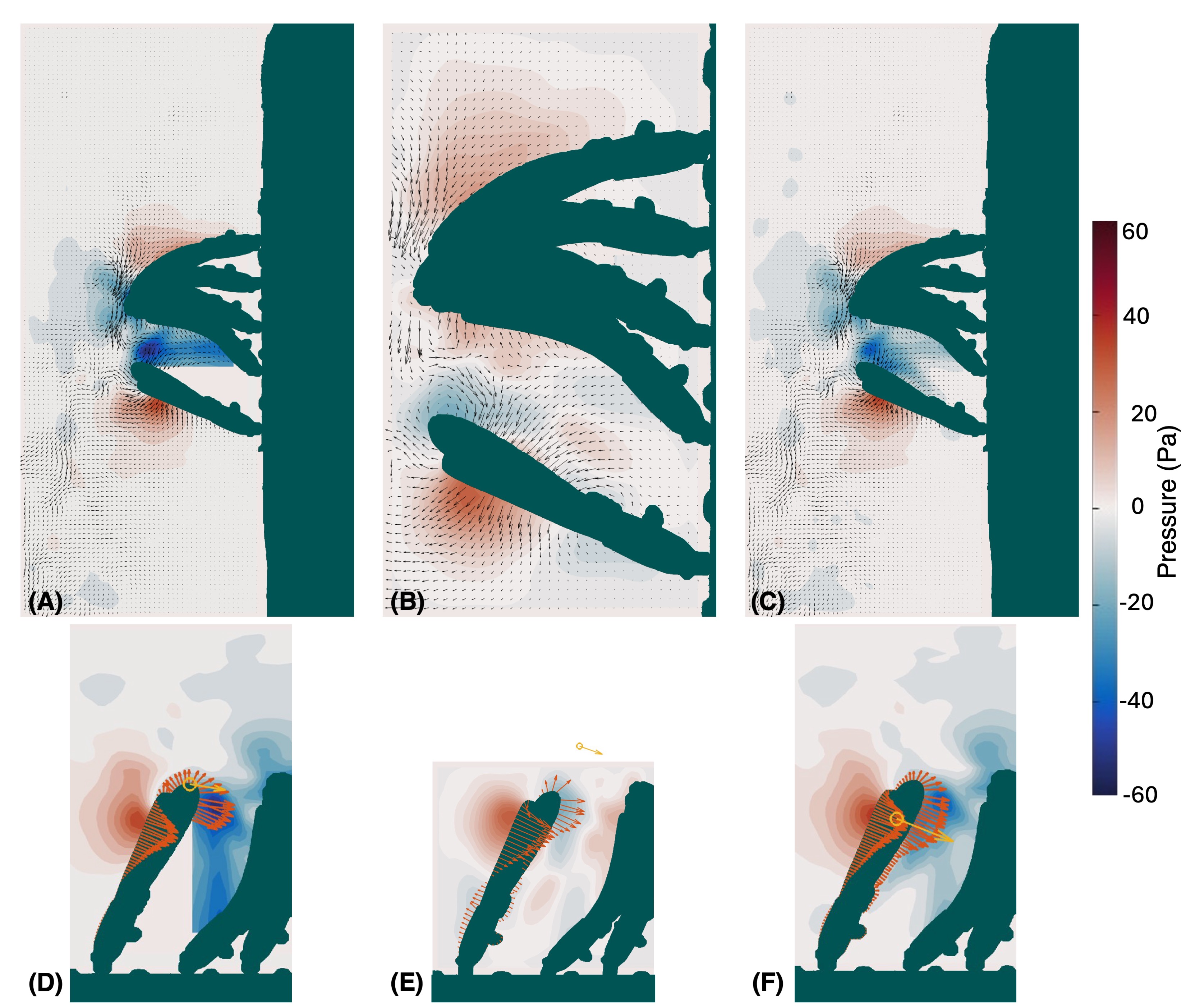}
\caption{Velocity fields and corresponding pressure calculations for the \emph{Nereus} shrimp-inspired robot performing metachronal rowing at a beat frequency of 1 Hz (see supplemental material, movie 1). (A) Laser-based PIV with a FoV of $193.16 \times 108.65 \ $mm. (B) Shadow-based PIV with a FoV of $86.23 \times 48.50 \ $mm. (C) Proposed dual-laser and shadow PIV technique with a FoV of $193.16 \times 108.65 \ $mm. Force distributions along the actuator surface calculated from (D) laser-based PIV, (E) shadow-based PIV, and (F) the new dual PIV technique. The yellow arrow and circle represent the resultant force and its application coordinate.}
\label{fig:Pressure}
\end{figure}

We recorded the metachronal beating of a shrimp-inspired underwater robot and applied our new technique (Figure \ref{fig:Pressure}). The robot has five actuators arranged in a row, each repeating a power and recovery stroke sequence with a phase lag relative to its neighbors, resulting in wave-like locomotion (Supplemental material, movie 1). While the laser-based data allow us to capture the entire robot within the FoV, shadows cast by the actuators cause the loss of about $3.64\%$ of the velocity field area (pure white regions in Figure \ref{fig:Pressure}A). Shadowed regions are in the immediate vicinity of the actuators, making them critical for the study of fluid-structure interactions. The shadow-based velocity field contains near-wall data, although its FoV is limited to the immediate neighborhood of the actuators (Figure \ref{fig:Pressure}B). These two simultaneous measurements enable us to reconstruct the complete velocity vector field for the robot, covering both far-field and near-wall velocities for the first time (Figure \ref{fig:Pressure}C).

When laser- and shadow-based techniques are used separately, missing data or a limited FoV can produce nonphysical results when deriving variables such as pressure. These types of calculations are sensitive to the FoV size, as longer integration paths yield better estimates. Despite the large FoV of the laser recordings, missing velocity values on the rear side of the bottom actuator result in an erroneous pressure field and no values near the real fluid-structure interface (Figure \ref{fig:Pressure}A). Shadow-based PIV overcomes the missing-data problem, enabling pressure computation that accounts for the actual fluid-structure interface (Figure \ref{fig:Pressure}B). However, the short integration paths within a FoV limited to the near vicinity of the actuators result in underestimated pressure values. Finally, using the full velocity field reconstructed from simultaneous laser and shadow recordings, we obtain a realistic pressure field (Figure \ref{fig:Pressure}C) that shows the positive and negative pressure regions expected from a rowing actuator during the power stroke \cite{colin_role_2020}. 

To illustrate the impact of different pressure calculations on a fluid-structure interaction analysis, we estimated the force distribution that the bottom paddle imparts to the fluid across three cases: laser PIV, shadow PIV, and dual laser and shadow PIV. We calculated the force imparted by the actuator to the fluid as

\begin{equation}
\overline{F} = \int \bm{\hat{n}} P \ \mathrm{d}A - \int \bm{\tau}\cdot \bm{\hat{n}} \ \mathrm{d}A
\end{equation}

where \bm{$\hat{n}$} is the outward-pointing unit normal vector at the actuator interface, $P$ is the fluid pressure, $\bm{\tau}$ is the viscous stress tensor, and $\overline{F}$ is the propulsion force imparted by the paddle to the fluid. Both terms in the equation are integrated over the actuator surface. Here, we neglect the shear term, assuming it is small relative to the pressure term. Given the similarity of the actuators to a flat plate largely perpendicular to the flow, the assumption is acceptable within the context of metachronal rowing \cite{herrera-amaya_propulsive_2024,colin_role_2020}. The force distribution obtained from the laser-based PIV data (Figure \ref{fig:Pressure}D) has a large portion of missing data, and the ``fake'' fluid-solid interface created by the laser shadow also causes the forces to be overestimated at the tip of the actuator. The resultant force, taking the base of the actuator as the moment center, is $F_{R}=0.51 \ N/m$ and is positioned at $103.8 \%$ of the actuator length. The shadow-based PIV data provide a complete force distribution; however, suction forces are overestimated on the front and underestimated on the rear side of the actuator. The differences in the pressure field result in a force of $F_{R}=0.35 \ N/m$ at $125.46 \%$ of the actuator length (Figure \ref{fig:Pressure}E). Finally, the force distribution from the dual technique shows both the frontal positive-pressure push and the rear negative-pressure suction on the actuator, leading to a calculation of  $F_{R}=0.86 \ N/m$ at $83.2 \%$ of the length (Figure \ref{fig:Pressure}F). The \emph{Nereus} shrimp robot actuator is flexible, just as metachronal animals in Nature \cite{ruszczyk_trends_2022}. Adequate force-distribution calculations are essential for studying and designing their propulsion dynamics.

\newpage

\section{Discussion}

The experimental study of FSI poses a significant challenge to traditional particle image velocimetry techniques, which were originally developed to measure velocities in unobstructed experimental volumes. The presence of solid objects in a laser-illuminated volume produces shadowed and light-scattering regions that lead to missing velocity information \cite{deparday_experimental_2022}. Many FSI problems allow for backlighting, making shadow-based PIV a laser-free alternative that can overcome these challenges in resolving flow near interfaces \cite{raffel_feasibility_2024}. Nonetheless, its application to macroscale problems has been limited by the technique's inherent FoV constraints. In this article, we showed that we can resolve both the far-field and near-wall velocity fields in FSI applications by simultaneously using laser- and shadow-based PIV recordings, rather than treating them as alternative techniques. 

Before applying the new technique to multiscale FSI applications, we performed the first simultaneous comparison of laser- and shadow-based PIV results. Additionally, we report one of the largest FoV in the literature for particle shadow velocimetry ($86.23 \times 48.50 \ $mm), closely followed by \cite{herrera-amaya_propulsive_2024}. At this scale, the seeding particle density converted the collimated LED backlight into diffuse illumination, increasing the noise-to-signal ratio. In our first experiment, we set the FoV equal for both recordings to compare performance at the macroscale. We showed that, with proper preprocessing techniques, shadow-based PIV can achieve close agreement with its laser counterpart. Both the high-pass filter and the tracer detection algorithm showed great agreement with the laser-based data, even when we increased the DoC by raising the numerical aperture of the lenses. The good agreement, even when the DoC was $185.6 \%$ larger than the laser thickness, indicates that even larger FoVs are achievable. The tracer detection algorithm showed slightly better vector-field similarity to the laser measurements (Figure \ref{fig:FieldmCMRI} \& \ref{fig:PCTCMRI}), but it is much more computationally expensive than the high-pass filter. The tracer detection algorithm and other more complex image-processing techniques for finding in-focus particles, such as inverse-problem approaches \cite{williams_inverse_2025}, can push the FoV of the shadow-based PIV even further if needed. 

Shadow-based PIV fields of view will never match those achievable with more traditional laser techniques. With our experimental setup, we can use the laser-recording camera to zoom out and capture a broad FoV, while using the shadow recordings to focus on the near-wall region, thereby obtaining a complete velocity field for FSI applications. In our second experiment, we set the zoom of the laser-recording camera to capture a FoV $400 \%$ larger than that of the shadow-based one. This allowed us to record the surroundings of the shrimp-inspired robot \emph{Nereus}, ensuring long integration paths for pressure calculations. Simultaneously, we used the shadow-based recording to capture the flow between the actuators, filling the missing data areas in the laser-based PIV. The reconstructed velocity field yielded better pressure estimates, thereby improving the force distribution calculations for the actuators --- a major improvement for the design of flexible underwater robotics. 

For simplicity, we interpolated the velocity values from the shadow-based grid to the laser grid, thereby maintaining a fixed grid spacing required by the QUEEN 2.0 pressure-calculation software. However, the shadow camera's higher zoom allows for more refined meshes than those from the laser velocity field. This enables higher spatial resolution near the solid interface, serving as the experimental analog to mesh refinement in computational fluid dynamics simulations. This dual technique is particularly useful for Boundary Layer (BL) experiments, where the shadow camera can resolve the BL thickness and the laser camera the far field. For example, our proposed methodology will improve data resolution in experiments aimed at understanding the role of millimeter-scale morphological changes in shark skin and their effects on overall swimming performance \cite{lauder_structure_2016} or in exploring the use of metamaterials to develop surfaces capable of controlling BL development \cite{avallone_metamaterials_2026}. The multiscale flow resolution will also be useful in the study of swarm fluid dynamics, allowing simultaneous measurement of organism-scale flows \cite{wilhelmus_observations_2014} and swarm-induced flows \cite{mohebbi_buoyancy-dependent_2026}.

Beyond FSI applications, the technique has the potential to measure multi-spatio-temporal turbulence scales, aiming to resolve dynamic ranges extending to the Kolmogorov scales \cite{clement_multi-spatio-temporal_2021}, and to enable bubble- and column-scale measurements of two-phase bubbly flow \cite{federle_interaction_2024, ravisankar_elastic_2025}. The technique is not without limitations, as the greater the separation between the spatial and temporal scales of the two recordings, the more difficult the measurements will be. There are many options for future users of the technique to achieve higher spatiotemporal resolution, for example, independently controlling camera recording speeds, implementing pulsed high-power illumination sources, using multiple tracer particle sizes, and combining PIV and PTV for velocity calculations. Going forward, we aim to expand the technique to stereo PIV (two dimensions and three components), using two cameras for the laser recordings \cite{prasad_stereoscopic_2000} and implementing general defocusing particle-tracking algorithms for the shadow-camera recordings \cite{barnkob_general_2015}. This work introduces a novel experimental methodology with the potential to push the frontiers of experimental fluid-structure interactions, providing the foundation for more accurate multiscale particle velocimetry measurements.

\section{Acknowledgements}
We acknowledge Dr. Margaret L. Byron at Penn State University for introducing Adrian Herrera-Amaya to shadow-based PIV techniques. We also gratefully acknowledge Nils Tack and Nina Mohebbi for their insightful discussions, which strengthened our work. We thank Marjorie Bradley for her assistance with coordination and administrative support throughout this research. Additionally, the authors acknowledge funding support from the Office of Naval Research, Bio-inspired Autonomous Systems Program (N000142412662).

\section{Author contributions}

A.H.A. conceptualization, formal analysis, methodology, writing --- original draft, review, and editing; M.M.W. conceptualization, formal analysis, methodology, supervision, funding acquisition, resources, project administration, writing--- review and editing.\\

The authors declare M.M.W. is the founder and technical lead of Nereus Systems, a company developing shrimp-inspired underwater robotic platforms. A.H.A. declares no competing interests.\\

\section{Data availability}

The data that support the findings of this study are available at the following URL : \url{https://drive.google.com/drive/folders/11RFqkd1GQNs_uw8L04HQx98i88ASUp_a?usp=sharing}
. Data will be posted in an open repository upon acceptance of the manuscript.

\newpage

\bibliography{references}% Produces the bibliography via BibTeX.

\end{document}